\documentclass[twocolumn,aps,prb,superscriptaddress,longbibliography]{revtex4-2}
\usepackage[utf8]{inputenc}
\usepackage{bm}
\usepackage{xcolor}
\usepackage{graphicx,amsmath}
\usepackage[colorlinks=true,linkcolor=blue,citecolor=blue]{hyperref}

\begin{document}

\title{Coexistence of Ferromagnetism and Antiferromagnetic Dimers \\ in Topological Insulators}
\author{Farhan Islam}
\affiliation{Ames National Laboratory, Ames, IA, 50011, USA}
\affiliation{Department of Physics and Astronomy, Iowa State University, Ames, IA, 50011, USA}

\author{Deborah Schlagel}
\affiliation{Ames National Laboratory, Ames, IA, 50011, USA}

\author{Yongbin Lee}
\affiliation{Ames National Laboratory, Ames, IA, 50011, USA}

\author{Santanu Pakhira}
\affiliation{Ames National Laboratory, Ames, IA, 50011, USA}

\author{Daniel M. Pajerowski}
\affiliation{Neutron Scattering Division, Oak Ridge National Laboratory, Oak Ridge, TN 37831, USA}

\author {David C. Johnston}
\affiliation{Ames National Laboratory, Ames, IA, 50011, USA}
\affiliation{Department of Physics and Astronomy, Iowa State University, Ames, IA, 50011, USA}

\author{Liqin Ke}
\affiliation{Ames National Laboratory, Ames, IA, 50011, USA}

\author{David Vaknin}
\affiliation{Ames National Laboratory, Ames, IA, 50011, USA}
\affiliation{Department of Physics and Astronomy, Iowa State University, Ames, IA, 50011, USA}

\author{Robert J. McQueeney}
\affiliation{Ames National Laboratory, Ames, IA, 50011, USA}
\affiliation{Department of Physics and Astronomy, Iowa State University, Ames, IA, 50011, USA}

\date{\today}

\begin{abstract}
    The addition of magnetic impurities in topological insulators can drive ferromagnetic order that leads to novel quantum anomalous Hall transport well below the Curie temperature.  The fragility of the quantized regime has been ascribed to the random nature of the magnetic moment distribution. Here, we refine this hypothesis by using inelastic neutron scattering and density-functional theory calculations to show that two antagonistic components define the magnetism in Mn-substituted SnTe, thereby limiting the effectiveness of dilute magnetic TIs. One component is strongly bound antiferromagnetic dimers that compete with ferromagnetic order. The other component consists of undimerized moments where ferromagnetic order develops via long-range interactions.
\end{abstract}

\maketitle

\section{Introduction}
The development of ferromagnetic (FM) order in topological insulators (TIs) is key to unlocking the quantum anomalous Hall effect (QAHE) where novel edge states can carry electrical current without dissipation~\cite{Qi2006,Qi2008,Liu2008,Yu2010,Nomura2011,Checkelsky2014}. The first successful approach to demonstrate the QAHE introduced ferromagnetism (FM) into non-magnetic and non-trivial topological insulators such as Bi$_2$Te$_3$ and Sb$_2$Te$_3$ through the substitution of low concentrations of magnetic ions~\cite{Zhang2009,Yu2010,Chang2013,Zhang2013,Zhang2013a,Chang2015}. Similar to dilute magnetic semiconductors, small concentrations of ions ($\leq5$\%) are capable of introducing FM order which presumably occurs through long-range interactions~\cite{Dietl2014,Dyck2002,Choi2004,Dyck2005,Hor2010,Watson2013,Zhang2013,Zimmermann2016,Islam2018,Vaknin2019,Teng2019,Yan2019b}. The random nature of magnetic substitution is hypothesized to introduce fundamental limitations to the temperature onset of the QAHE, which motivated the search for intrinsic magnetic topological insulators, leading to the discovery of MnBi$_2$Te$_4$~\cite{Otrokov2019,Li2019,Zhang2019,Liu2020,Chen2019,Yan2019,Gong2019,Deng2020,Hao2019,Lee2019,Xu2022,Zhang2022}.  However, detailed understanding of the magnetic interactions in the dilute case and why QAHE is suppressed remains poorly understood~\cite{Kou2013,Tokura2019}. One may even question how long-range magnetic interactions, necessary for FM order in dilute systems, can be mediated without a high density of conduction electrons.
\begin{figure}[!h]
	\centering
	\includegraphics[width=1\linewidth]{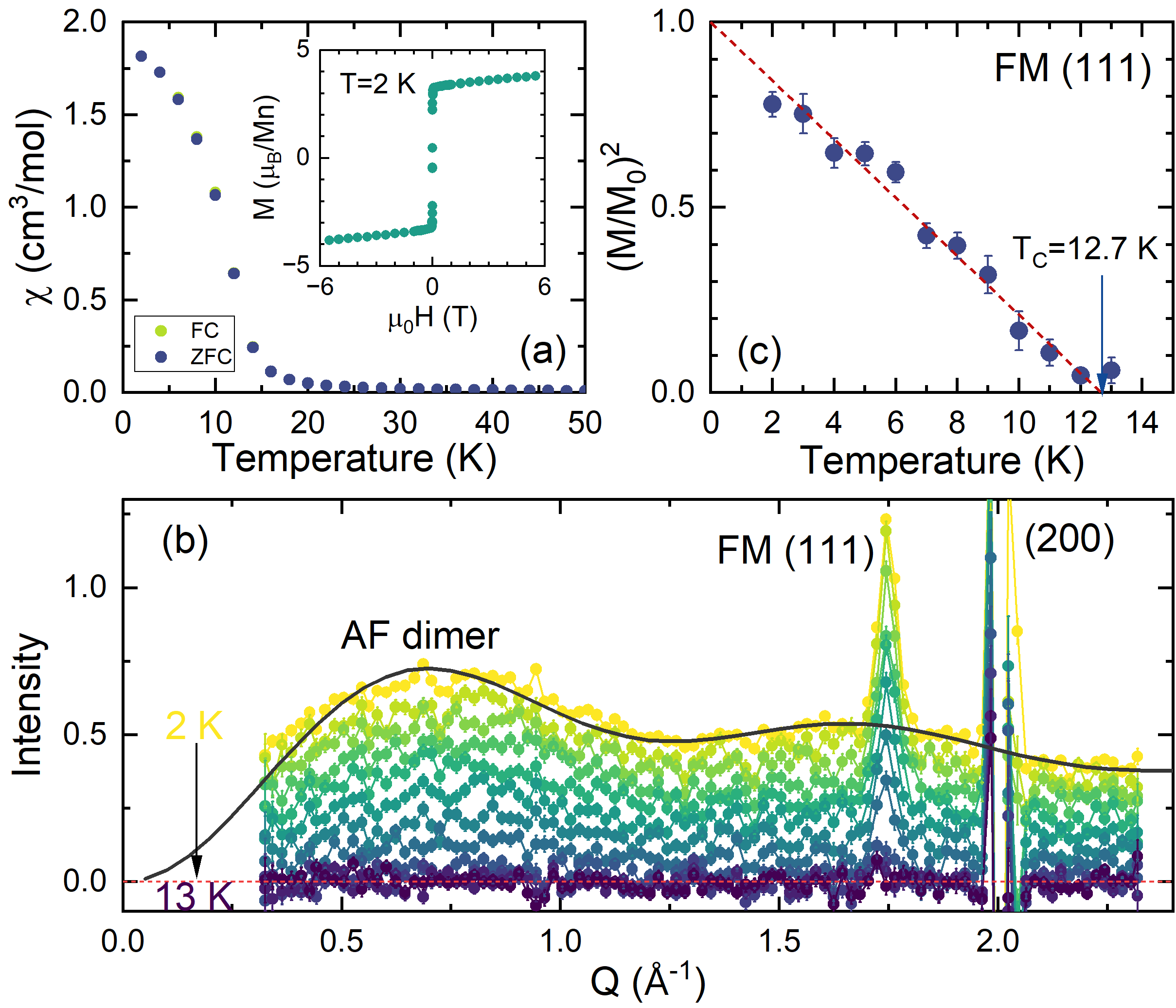}
	\caption{ \textbf{Long-range ferromagnetic order in Sn$_{0.9}$Mn$_{0.1}$Te.} (a) Field-cooled (FC) and zero-field cooled (ZFC) magnetic susceptibility measured in a magnetic field $H=$ 0.1 T showing a smeared onset of ferromagnetic order near 15 K. The inset shows the low-temperature magnetization versus field that is characteristic of a soft ferromagnet. (b) Elastic neutron-scattering intensity (integrated from $E=-0.05$ to 0.05 meV) as a function of momentum transfer plotted for several temperatures after subtracting 20 K data. The magnetic component of the (111) reflection is indicated, as well as broad diffuse scattering consistent with NNN antiferromagnetic dimers. The data are not offset. (c) Ferromagnetic order parameter obtained from the integrated intensity of the magnetic (111) reflection. }    
    \label{Fig:characterization}
\end{figure}

\begin{figure*}[]
	\centering
	\includegraphics[width=0.9\linewidth]{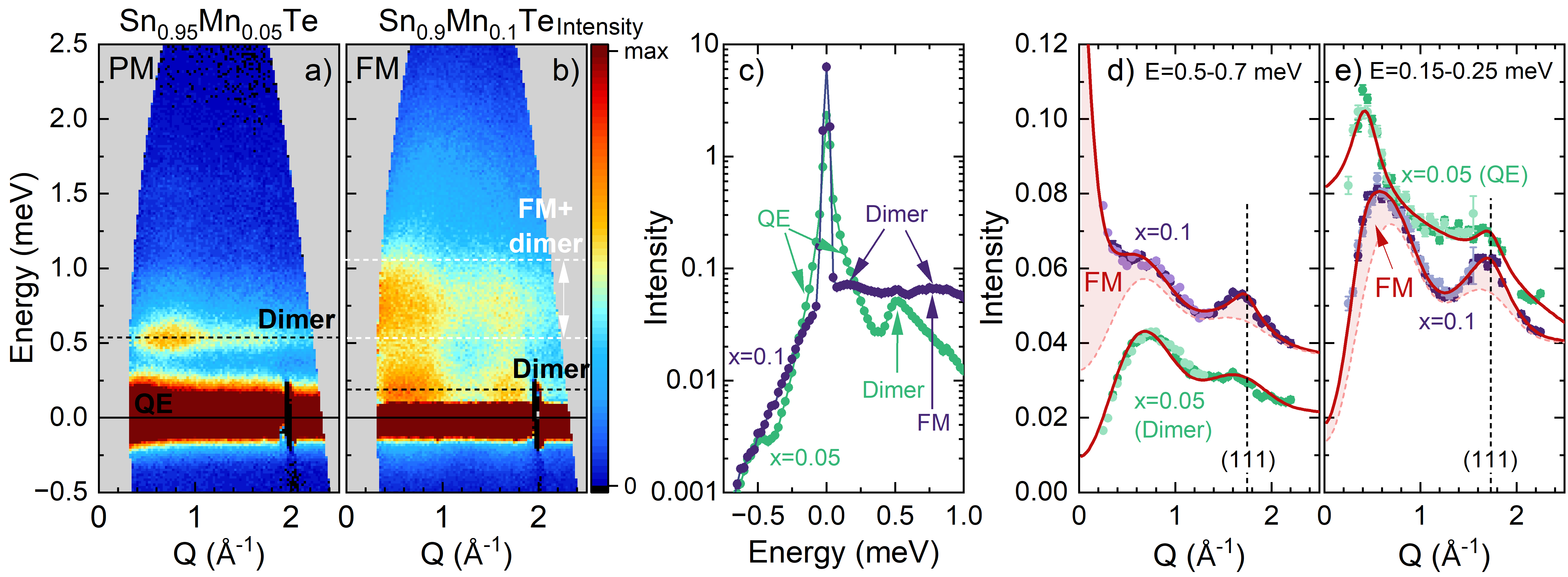}
	\caption{ \textbf{Low-temperature magnetic excitations in Sn$_{1-x}$Mn$_{x}$Te.} Powder inelastic-neutron-scattering spectra measured at $T=2$ K and $E_i=3.32$ meV for (a) the paramagnetic $x=0.05$ sample and (b) the ordered ferromagnetic $x=0.1$ sample.  Horizontal lines or regions highlight quasielastic (QE), dimer, and FM resonance excitations. (c) Energy spectra of $x=0.05$ (green) and 0.1 (blue) samples measured with $E_i=1.55$ meV and averaged over the full $Q$ range of panels (a) and (b).  Arrows highlight spectral features seen in panels (a) and (b). $Q$-dependence of magnetic spectra of the $x=0.05$ (green) and $x=0.1$ (purple) samples at (d) $E=0.5$-0.7 meV and (e) $E=0.15$-0.25 meV.  In (d) and (e) light (dark) symbols are measured with $E_i=1.55$ (3.32) meV, respectively. Solid red lines are fits to dimer plus ferromagnetic lineshapes, as described in the Supplementary Information.  The dashed pink line shows the dimer contribution of the $x=0.1$ sample and the shaded region is the ferromagnetic contribution.}     
    \label{Fig:INSoverview}
\end{figure*}
Recent inelastic-neutron-scattering (INS) measurements raise additional questions by finding strong antiferromagnetic (AF) interactions in paramagnetic (PM) Sb$_2$Te$_3$~\cite{Islam2023} and SnTe~\cite{Vaknin2020} topological insulators with low concentrations of Mn substitution. AF spin dimers form between next-nearest-neighbor (NNN) Mn atoms possessing Mn-Te-Mn linear bonds, following the well-known Goodenough-Kanamori rules~\cite{Goodenough1963}. The AF interactions are found to be many times stronger than any FM couplings and form tightly-bound dimer-singlet states. This suggests the existence of a two-component magnetic system where AF dimer singlets coexist with weakly-coupled and undimerized Mn ions. It is an interesting and fundamental problem to understand how FM order develops when the Mn concentration is increased beyond the threshold for FM order.

SnTe is one of the first known examples of a topological crystalline insulator~\cite{Hasan2010,Qi2011,Hsieh2012,Tanaka2012,Xu2012,Dziawa2012}.  In this system, surface electronic states are protected by time-reversal and crystalline symmetries. SnTe  adopts a cubic rocksalt structure~\cite{BauerPereira2013} and FM order with the substitution of Mn for Sn~\cite{Inoue1977,Vennix1993,Eggenkamp1994}.  FM order develops above a critical composition of about 5\% where the exact compositional crossover depends on thermal treatment~\cite{Vaknin2020}. The underlying mechanism responsible for the development of ferromagnetic order is still debated.

Here, we extend the study to 10\% Mn-doped SnTe, and for completeness, we extensively compare the two systems. Our study aims to unravel the characteristics of the magnetic interactions that drive the development of ferromagnetism in this 10\% Mn-doped SnTe. AF dimer excitations persist at these compositions and are found to be split by the average molecular-field of FM-ordered Mn ions. We also observe a FM resonance which provides a direct measure of the molecular field strength and tracks the FM order parameter. These studies enforce the two-component hypothesis in the PM and FM regimes for both SnTe and (Bi,Sb)$_2$Te$_3$ dilute magnetic TI families. In the two-component model, undimerized Mn ions are coupled by weak, but long-ranged, magnetic interactions that form a backbone of the FM order. The strong AF dimers are weakly polarized through interaction with the molecular field of the undimerized spins, which limits the global magnetization and introduces spatial inhomogeneities.

The formation of Mn-Mn AF dimers may explain why the QAHE has only been observed for Cr- and V- substituted TIs, but never with Mn substitutions. To investigate this point, we conducted density-functional-theory (DFT) calculations of the pairwise magnetic interactions between two $M$ ions in SnTe where $M=$ V, Cr, Mn. In the case of Mn, DFT finds that the NNN dimer is strongly AF, in agreement with experiment, but fails to reproduce the additional FM interactions between other pairs that are needed for the development of long-range FM order. Similar to Mn, the NNN V-V dimer also exhibits a weaker AF dimer coupling, but nearest-neighbor (NN) and other couplings favor ferromagnetism. Surprisingly, all couplings with Cr doping are FM with NNN Cr-Cr dimer exhibiting very strong FM coupling, consistent with experimental reports of high-temperature FM order in Cr-substituted SnTe~\cite{Inoue1981, Wang2018, Muhammad2022}. These results suggest that Cr and V substitutions are more favorable than Mn to host robust topological states with broken time-reversal symmetry.
 
\begin{figure*}[]
	\centering
	\includegraphics[width=1\linewidth]{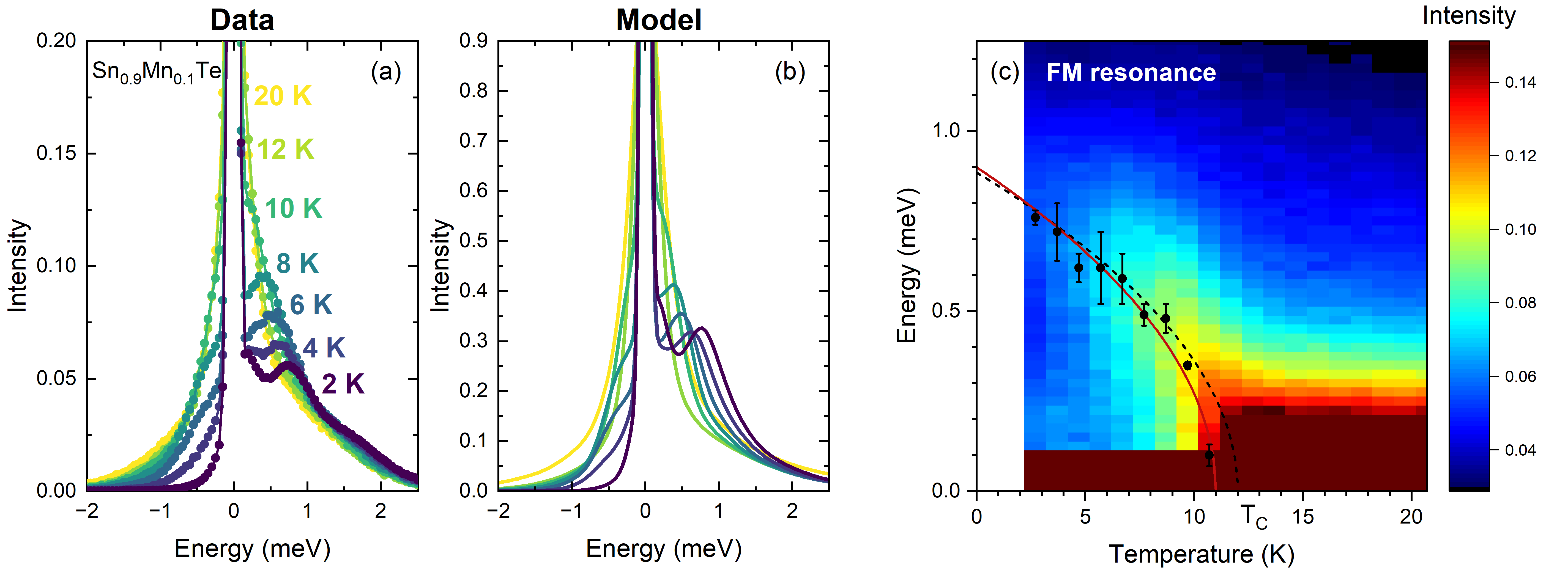}
	\caption{ \textbf{Ferromagnetic resonance excitation}. (a) The magnetic energy spectrum averaged over $Q=$ 0.7$-$1.5~\AA~for several temperatures showing the evolution of the FM resonance peak. (b) Calculations of the magnetic spectrum using the two-component magnetism model. (c) Image plot of the magnetic spectrum data from panel (a) for all temperatures.  A fit (dashed line) to the resonance peak (circles) is shown together with a fit (red line) to a mean-field-like order parameter with $T_{\rm C}=11$ K and $\Delta_0=0.9$ meV as described in the text. The dashed line is the mean-field order parameter with $T_{\rm C}=12$ K and $\Delta_0 = 3k_{\rm B}T_{\rm C}/(s+1)$.}     
    \label{Fig:resonance}
\end{figure*}

\section{Methods}
\subsection{Crystal Growth and Characterization}
Polycrystalline samples of Sn$_{1-x}$Mn$_x$Te (where $x=0$, 0.05, and 0.1) were synthesized using solid-state reaction from stoichiometric quantities of Sn, Mn, and Te. Analysis of scanning electron microscopy and energy-dispersive spectroscopy show the $x=0.1$ sample to be single phase with a composition of $x=0.10(1)$ and X-ray powder diffraction measurements confirm the SnTe structure. Magnetization measurements were carried out using a Quantum Design MPMS magnetometer. For more details, see the Supplementary Information (SI).

\subsection{Inelastic Neutron Scattering}
Inelastic neutron scattering measurements were performed on all three samples of Sn$_{1-x}$Mn$_x$ with $x=$~0,~0.05,~and~0.1 using the CNCS spectrometer at the Spallation Neutron Source at Oak Ridge National Laboratory.  Incident neutron energies of $E_{\rm i} = 1.55$ and 3.32 meV were chosen and the samples were measured at several temperatures from 2 to 20 K. The INS intensity is reported as $I(Q,E)$ which is proportional to the spin-spin correlation function between Mn ions.

\section{Experimental data}

The Sn$_{1-x}$Mn$_x$ sample remains paramagnetic for $x=0.05$ and displays no long-range FM order, as reported previously~\cite{Vaknin2020}. Conversely, the $x=0.1$ sample does display FM order with a Curie temperature $T_{\rm C} \approx 12$ K, as observed from magnetic susceptibility and magnetization measurements shown in Fig.~\ref{Fig:characterization}(a). Other details of sample synthesis and characterization are given in the Supplementary Information (SI).

We performed inelastic neutron scattering experiments on powders of Sn$_{1-x}$Mn$_x$Te with $x=$ 0, 0.05, and 0.1 using the Cold Neutron Chopper Spectrometer at the Spallation Neutron Source.  Measurements were performed at several temperatures with incident neutron energies of $E_i=$1.55 and 3.32 meV. Figure~\ref{Fig:characterization}(b) shows elastic-neutron-scattering data from the $x=0.1$ sample at several temperatures after subtraction of the $T=20$ K data.  There is clear evidence for long-range FM ordering that contributes to the (111) reflection where the nuclear contribution is very weak (see SI).  The integrated intensity of the FM (111) peak is shown in Fig.~\ref{Fig:characterization}(c) and displays mean-field-like evolution with an estimated $T_{\rm C}=12.7 (7)$~K.
    
As reported previously, INS data from the paramagnetic $x=0.05$ compositions contain strongly-coupled AF dimers with exchange coupling of $\mathcal{J}_2\approx -0.58$~meV~\cite{Vaknin2020}. Figures~\ref{Fig:INSoverview}(a) and \ref{Fig:INSoverview}(c) display INS data for $x=0.05$ where the principal feature is a dimer spin-state transition from the $S=0$ singlet ground state to the $S=1$ triplet excited state with energy $E = |\mathcal{J}_2|$. The excitation has a characteristic $Q$-dependent dimer structure factor, as shown in Fig.~\ref{Fig:INSoverview}(d). The dimers consist of NNN Mn ions where linear Mn-Te-Mn bonds mediate strong AF superexchange coupling.  At energies below 0.2 meV, Figs.~\ref{Fig:INSoverview}(a) and \ref{Fig:INSoverview}(c) indicate that strong quasielastic magnetic fluctuations are present, as expected for a paramagnet that is close to FM order. Figure~\ref{Fig:INSoverview}(e) shows that the quasielastic magnetic fluctuations are strongly enhanced at small $Q$ for $x=0.05$ which is a signature of FM character.

The spectrum of the FM-ordered $x=0.1$ sample is shown in Figs.~\ref{Fig:INSoverview}(b) and \ref{Fig:INSoverview}(c). The quasielastic fluctuations are absent and the inelastic spectrum now consists of contributions from AF-dimers and FM excitations.  AF dimers survive in the FM-ordered state, but their spectral weight is smeared out and a strong component shifts to low energies ($<0.25$ meV). Analysis of the structure factor below 0.25 meV in Fig.~\ref{Fig:INSoverview}(e) shows a dominant dimer contribution. Dimer scattering also appears in the elastic scans shown in Fig.~\ref{Fig:characterization}(b), which could indicate that static AF (spin glass) correlations are present \cite{Vennix1993}. Broad spectral features between $0.5$ and $1.0$ meV in Figs.~\ref{Fig:INSoverview}(b) and \ref{Fig:INSoverview}(d) consist of overlapping contributions from dimers and FM-ordered Mn ions. The FM correlations in this energy range increase sharply as $Q\rightarrow0$ and also form a weaker peak at the (111) Brillouin zone center.

 Next, we describe a characteristic excitation of the long-range-ordered dilute magnet, identified as a FM resonance. Deep within the ordered state, the FM resonance contributes to a broad, low-temperature peak at 0.8 meV as described above, and shown in Fig.~\ref{Fig:resonance}(a). As the temperature is increased, Figs.~\ref{Fig:resonance}(a) and \ref{Fig:resonance}(c) show that the FM resonance shifts to lower energies and forms a quasielastic feature above $T_{\rm C}$.  The fitted values of the FM resonance energy ($\Delta$) evolve according to a mean-field-like magnetic order parameter ($\Delta = \Delta_0 \sqrt{1-T/T_{\rm C}}$). As described below, the resonance energy is associated with the Zeeman splitting of the Mn spin states in the self-consistent molecular field of surrounding Mn ions. 

\begin{figure*}[]
	\centering
	\includegraphics[width=1\linewidth]{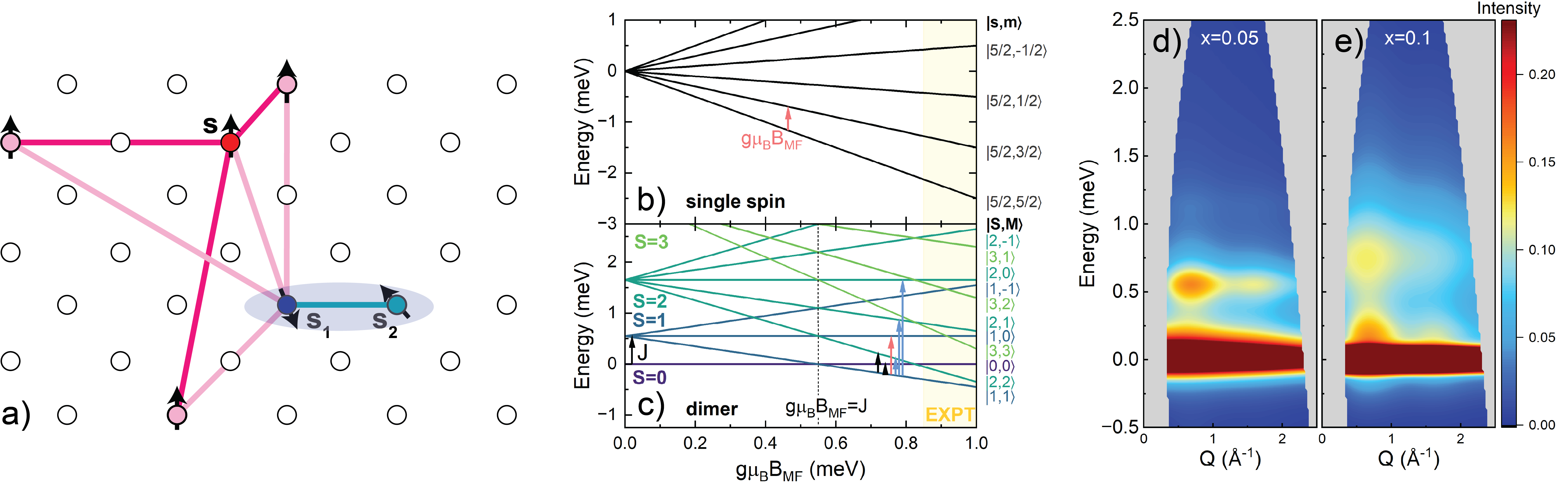}
	\caption{ \textbf{Two-component magnetic model}. (a) Depiction of the interactions of an undimerized Mn spin (red circle) with other undimerized spins (pink circles) and interactions of a dimerized spin (blue circle) with other dimerized (cyan) and undimerized spins.  (b) Level diagram for a single $s=5/2$ spin in the molecular field ($B_{\rm MF}$) of other undimerized spins.  (c) The level diagram of an antiferromagnetic dimer with exchange coupling $\mathcal{J}_2$ in a molecular field. Vertical arrows show dipole-allowed $S=0$ to 1 (black) and $S=1$ to 2 (blue) dimer spin state transitions and $\Delta S=0$ or $\Delta s = 0$ Zeeman transitions (red) out of the ground state.  The vertical dashed line in (c) marks level crossing into the $S=1$ dimer ground state and the yellow area highlights the approximate experimental value of the ground state molecular field strength. Two-component model calculations of the inelastic neutron scattering spectrum at $T=2.2$~K for (d) $x=0.05$ and (e) $x=0.1$ using the parameters in Table \ref{tab:params}.}  
    \label{Fig:levels}
\end{figure*}

\section{Two-component magnetism model}
Mn-doped SnTe and other dilute magnetic topological insulators are unusual ferromagnets because the strongest interaction is AF.  Here, we describe a simple two-component magnetic model that differentiates between undimerized Mn ions (spins with no NNN neighbors), those with one NNN neighbor (a dimer), and those with more than one NNN neighbor (trimers and larger AF clusters).

{\it Random magnetic alloy}. For the random substitution of $xN$ Mn ions onto $N$ Sn sites, the fraction of undimerized Mn ions is $f_{\rm u} = (1-x)^6 \approx 53\%$ (the total number of undimerized Mn ions $N_{\rm u}=Nxf_{\rm u}$).  The fraction of Mn ions that participate in a NNN dimer is $f_{\rm d}=6x(1-x)^{10} \approx 21\%$, forming $N_{\rm d}=Nxf_{\rm d}/2$ total dimers (ignoring NN occupations) for $x=0.1$.  All other Mn ions ($N_{\rm c}=Nx-N_{\rm u}-2N_{\rm d}$) are part of larger NNN clusters which have a more complicated spectrum that is assumed to form a background-scattering. See Table \ref{tab:params} for more details.

FM order can only develop in dilute systems when the concentration of magnetic ions exceeds the percolation threshold. For an FCC sublattice with NN FM interactions, the site percolation threshold of $x_{\rm p} = 0.198$ is insufficient to establish order. Longer-range interactions are needed.  For example, $x_{\rm p}\approx 0.056$ for interactions up to 4 nearest neighbors (excluding the NNN shell)~\cite{Gawron1991} which is easily exceeded in the $x=0.1$ sample.

 {\it Two-component Hamiltonian}. We hypothesize that the undimerized spins are coupled by long-range interactions that form a FM-ordered backbone when the Mn concentration exceeds the percolation threshold. AF NNN dimers are only weakly polarized by coupling to the FM-ordered spins.

Within a mean-field approximation, the undimerized and dimerized Hamiltonians become decoupled (see SI for a derivation) and are given by
\begin{align}
    \mathcal{H}_{\rm u} &= -g\mu_{\rm B} B_{\rm MF}s^z \label{Eq:Hamiltonian1} \\
    \mathcal{H}_{\rm d} &= -\mathcal{J}_2{\bf s}_1\cdot {\bf s}_2 + \mathcal{H}_{\rm u,1} +  \mathcal{H}_{\rm u,2} \label{Eq:Hamiltonian2}
\end{align}
where $B_{\rm MF}$ is the molecular field generated by the undimerized spins and $\mathcal{J}_2= -0.58$~meV~\cite{Vaknin2020} is the AF coupling between NNN dimers. In Eq.~(\ref{Eq:Hamiltonian2}), $\mathcal{H}_{\rm u,i}$ is the interaction of the molecular field with spin $i$ in the NNN dimer [Eq.~(\ref{Eq:Hamiltonian1})].  The molecular field is a sum over coordination shells ($k$) with FCC coordination number $z_k$ which excludes the NNN ($k=2$) shell, given by
\begin{equation}
g \mu_{\rm B} B_{\rm MF} =  xs\sum_{k \neq 2} z_k\mathcal{J}_k.
\end{equation}

The average molecular field in the ground state can be estimated from the Curie temperature $T_{\rm C} = s(s+1)(x\sum_{k \neq 2} z_k \mathcal{J}_k)/3k_{\rm B} \approx 12.7$ K, resulting in a value of $g\mu_{\rm B} B_{\rm MF}(0) = 3k_{\rm B} T_{\rm C}/(s+1)\approx$ 0.95 meV.  Within mean-field approximation, the molecular field is $B_{\rm MF}=0$ above $T_{\rm C}$ and
\begin{equation}
    g\mu_{\rm B} B_{\rm MF}(T) = g\mu_{\rm B} B_{\rm MF}(0)\sqrt{1-\frac{T}{T_{\rm C}}}
\end{equation}
below $T_{\rm C}$. Figures~\ref{Fig:characterization}(b) and \ref{Fig:resonance}(c) show that both the order parameter and FM resonance follow a mean-field form and we can associate $\Delta_0=g\mu_bB_MF(0)$ and $\Delta = g\mu_bB_MF(T)$.

For the undimerized spin $s$, the eigenstates of $\mathcal{H}_{\rm u}$ are the conventional $|s,m\rangle$ spin states projected along the molecular field direction (chosen to be the $z$-axis) and with Zeeman energy $E_m=-g\mu_{\rm B}B_{\rm MF}m$. The level spectrum for $s=5/2$ is shown in Fig.~\ref{Fig:levels}(b).

For the AF dimer with $B_{\rm MF}=0$, the energy levels depend only on the total spin $\mathcal{S}=0,...,2s$ of the dimer $E_\mathcal{S}  = -\frac{\mathcal{J}_2}{2}[\mathcal{S}(\mathcal{S}+1) - 2s(s+1)]$ and each level has a degeneracy of $2\mathcal{S}+1$. The dimer spin states can be written in the $m_1,m_2$ basis as $|\mathcal{S},\mathcal{M}\rangle = \sum_{m_1m_2}c_{m_1m_2}^{\mathcal{S}\mathcal{M}}|s,m_1\rangle|s,m_2\rangle$, where $c_{m_1m_2}^{\mathcal{S}\mathcal{M}}$ is a Clebsch-Gordan coefficient. 
The molecular field acting on the dimer will result in a Zeeman splitting of the dimer energy levels, but $\mathcal{M}$ remains a good quantum number, as shown in Fig.~\ref{Fig:levels}(b).

\section{Data Analysis}
Using the Hamiltonians (\ref{Eq:Hamiltonian1}) and (\ref{Eq:Hamiltonian2}), the INS intensity of the two-component model is given by
\begin{equation}
    I(Q,E) = N_{\rm u}I_{\rm u}(Q,E)+N_{\rm d}I_{\rm d}(Q,E)+N_{\rm c}I_{\rm c}   
\end{equation}
where the intensity from larger AF clusters ($I_{\rm c}$) are assumed to form a featureless background and is not considered further. The explicit forms for $I_{\rm u}$ and $I_{\rm d}$ are given in the SI.

Figures~\ref{Fig:levels}(d) and \ref{Fig:levels}(e) show calculations of the low-temperature spectra for $x=0.05$ and $x=0.1$ samples with the parameters given in Table \ref{tab:params}. This simple approximation does not include spatial correlations of the FM-ordered ions, and is of limited value in analyzing the $Q$-dependence of the excitations.  However, Figs. \ref{Fig:resonance} and \ref{Fig:levels} show that the model reproduces the spectral features of both samples with only three parameters: the Mn concentration ($x$), the AF dimer exchange ($\mathcal{J}_2$), and the Curie temperature ($T_{\rm C}$) where the latter parameter determines the strength and temperature dependence of the molecular field. We also include parameters, i.e. $\gamma$ and $\Gamma$ in Table~\ref{tab:params}, that represent spectral broadening (see SI).

The two-component model associates the FM resonance energy with the Zeeman splitting of the undimerized Mn ions in the molecular field.  As neutron scattering observes transitions with $\Delta m = \pm 1$, the resonance measures directly the average molecular field at $T<T_{\rm C}$ ($\Delta = g\mu_{\rm B}B_{\rm MF}$), as shown in Fig.~\ref{Fig:levels}(b). Figure~\ref{Fig:resonance} confirms the resonance energy $g\mu_{\rm B} B_{\rm MF} \approx 0.95$ meV and the temperature dependence of the FM resonance follows the two-component mean-field model.

Dimer excited states are also split by the molecular field and can contribute to the FM resonance intensity [red arrow in Fig.~\ref{Fig:levels}(c)].  The dimer level diagram also indicates that the molecular field is sufficiently strong at low temperatures to induce a level crossing to $|\mathcal{S},\mathcal{M}\rangle = |1,1\rangle$ ground state.  This level crossing creates intense dimer excitations at low energies, corresponding to $|1,1\rangle\rightarrow |0,0\rangle$ and $|1,1\rangle\rightarrow |2,2\rangle$ spin-state transitions [black arrows in Fig.~\ref{Fig:levels}(c)].  Higher-energy dimer excitations [blue arrows in  Fig.~\ref{Fig:levels}(c), such as $|1,1\rangle\rightarrow |2,0\rangle$] create a broad spectrum of dimer excitations up to 1 meV or more.

\begin{table}[]
    \centering
\begin{tabular}{ c | c | c | c | c |c | c  }
\hline\hline
 $ x$ & $f_{\rm u}$ & $f_{\rm d}$ & $\mathcal{J}_2$ (meV) 	& $T_{\rm C}$ (K) 	&   $\gamma$ (meV) & $\Gamma$ (meV)	\\
\hline
0.05			&  0.735 			  &  0.180			& -0.58			& 0		& 0.28		& 0.07 \\
0.10			&  0.531 			  &  0.209			& -0.58			& 12.7	& 0.85		& 0.28 \\
\hline\hline
    \end{tabular}
    \caption{Parameters of the two-component model. $x$ is the concentration of Mn, $f_{\rm u}$ ($f_{\rm d}$) is the undimerized (dimerized) fraction of Mn, $\mathcal{J}_2$ is the AF dimer exchange, $T_{\rm C}$ is the Curie temperature, and $\gamma$ ($\Gamma$) is the inelastic (quasielastic) width.}
    \label{tab:params}
\end{table}

\section{First-principles calculations}
Dimer singlet formation is surprisingly ubiquitous in all Mn and Te-based topological insulators.  Strongly-coupled AF Mn-Te-Mn dimers form in Mn-substituted Sb$_2$Te$_3$~\cite{Islam2023} and are known to form in MnBi$_2$Te$_4$ and MnSb$_2$Te$_4$ intrinsic AF-TI when anti-site mixing defects are present \cite{Lai2021}. In each case, singlet formation is tied to the presence of linear Mn-Te-Mn bonds that foster strong AF superexchange. However, one caveat is that there are no known observations of the QAHE in Mn-substituted TIs, suggesting that AF dimer formation is deleterious. On the other hand, Cr- and V-substituted TIs have demonstrated QAHE, so it is reasonable to ask whether AF dimer formation can also be found in these compounds. Although our experimental and theoretical work focuses on Mn, we conduct DFT calculations to also evaluate the pairwise exchange couplings between two $M$ ions in SnTe where  $M=$ Mn, V, or Cr. 

Our calculations are based on constructing a 3$\times$3$\times$3 SnTe supercell composed of 216 atoms, with a pair of Sn sites substituted by two $3d$ transition-metal ($M$) ions to form a dimer. Nine configurations of $M$-$M$ dimers with different bond lengths are considered by substituting on different Sn sites.  The resulting $M$ concentration of our model $<1\%$ is much lower than in experimental samples, and we disregard the possible influence of nearby impurities on the magnetic interaction within one dimer, which may be present in real samples.  For each dimer, the energies of FM ($E_\text{FM}$) and AFM ($E_\text{AFM}$) configurations were calculated to estimate the dimer exchange coupling.  Further details of the DFT calculations can be found in the SI.
\begin{figure}[htb]
  \centering
  \begin{tabular}{cc}
    \includegraphics[width=0.50\linewidth,clip] {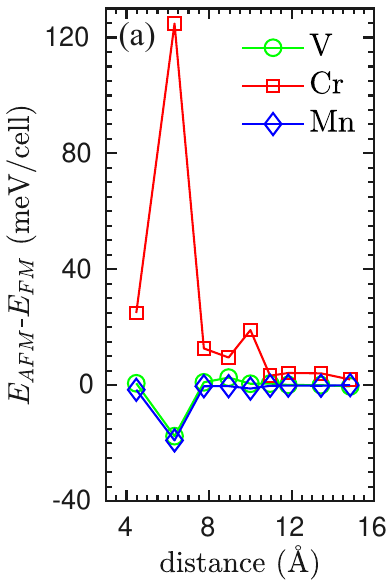}  &  
    \includegraphics[width=0.50\linewidth,clip] {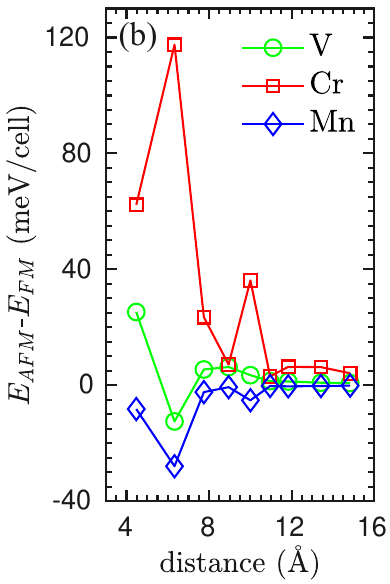}
  \end{tabular}%
  \caption{Dimer magnetic energy as a function of dimer length calculated using the (a) non-relaxed and (b) relaxed crystal structures.
The magnetic energy is defined as the difference between the AFM and FM configurations, $E_\text{AFM} - E_\text{FM}$.
For Mn dimers and V dimers without structural relaxation, only the NNN dimer shows a similarly large AFM coupling, while all other dimer configurations show negligible magnetic coupling.
In contrast, Cr dimers exhibit an overall large FM coupling, especially for the NNN dimer.}
  \label{fig:magnetic-energy-relaxed}
\end{figure}

Figures~\ref{fig:magnetic-energy-relaxed}(a) and \ref{fig:magnetic-energy-relaxed}(b) show the magnetic energy of the dimers, calculated using non-relaxed and relaxed crystal structures, respectively. Both show a prominent minimum at the NNN position for Mn, confirming the strong AF NNN coupling observed experimentally. Notably, without relaxation, V and Mn dimers exhibit nearly identical behavior, demonstrating a significant AF coupling for the NNN dimer, while all other dimer configurations show negligible coupling.
In contrast, the interaction between NNN Cr differs markedly from that of V and Mn counterparts, displaying a strong overall FM coupling.  This remarkable difference can be traced back to the $e_{\rm g}$-orbital filling and degeneracies, as explained in the SI.
Relaxation of the structure affects the magnetic coupling between two $M$ atoms for the shorter-distance dimer configurations, as shown in Figs. \ref{fig:magnetic-energy-relaxed}(a) and \ref{fig:magnetic-energy-relaxed}(b).
The AFM coupling for the NNN V-V dimer becomes weaker than that of the Mn-Mn dimer.
Overall, the dominant AFM interaction for the NNN Mn-Mn dimer aligns with experiments and previous observations~\cite{Vaknin2020,Islam2023}.
The FM coupling of Cr is also consistent with previous experiments for both bulk~\cite{Inoue1981, Muhammad2022} and film~\cite{Wang2018} samples, where Cr-doped SnTe is found to be FM with a high Curie temperature. Further evaluation of the evolution of magnetic interactions using the band-filling model~\cite{Ke2015} is provided in the SI.

\section{Conclusion}
Mn-doped SnTe develops long-range FM order despite the fact that the largest pairwise magnetic interactions are AF.  The maximum magnetic interaction energy of the NNN coordination shell is 
$6\mathcal{J}_2=-3.5$ meV, whereas all other shells have $\sum_{k \neq 2} z_k\mathcal{J}_k = 3.8$ meV. Therefore, the overall interaction slightly favors ferromagnetism within a mean-field treatment. However, $T_{\rm C}$ is determined only by the FM exchange couplings, not the total (which would result in $T_{\rm C} \approx 1$ K). This justifies the essential premise of the two-component model where the formation of NNN AF dimer singlets only weakly perturbs the devemopment of FM order.

In the two-component model, although $T_{\rm C}$ is weakly affected by dimer formation, the overall magnetization is suppressed as a result of dimer formation. The level crossing of the dimer in the molecular field of the undimerized spins (into the $|1,1\rangle$ state) results in a weak magnetization of $g\mu_{\rm B}\langle \mathcal{M}\rangle = 2~\mu_{\rm B}$ per Mn-Mn dimer. The magnetization per Mn expected from the two-component model at $T=0$ is $M\approx(5~\mu_{\rm B})f_{\rm u} + (2~\mu_{\rm B})f_{\rm d}/2 = 2.8~\mu_{\rm B}$, which is close to the measured value of 3.2 $\mu_{\rm B}$ in Fig.~\ref{Fig:characterization}(a).

Despite the success of the two-component model, it is clear that it does not include spatial correlations which are complicated by the presence of both competing magnetic interactions and chemical disorder.  Thus, the model is incapable of describing any short-range order, such as spin-glass or cluster formation.  Such disorder has been reported previously~\cite{Vennix1993} and is hinted upon in the large quasi-static diffuse scattering observed in the elastic channel [Fig.~\ref{Fig:characterization}(b)].  The two-component model also ignores the collective character of FM excitations which are expected to form damped magnons as observed in Mn-substituted Bi$_2$Te$_3$~\cite{Vaknin2019}. Future measurements on single-crystal samples should be able to address these key details.

DFT calculations confirm NNN AFM Mn-Mn coupling, consistent with our experimental results. While strong AF interactions may prevent the development of the QAHE in Mn-substituted samples, the results from DFT calculations suggest that AF dimer singlets form in V-substituted topological insulators, where the QAHE has been observed. However, DFT calculations also suggest that V-substituted samples have larger overall FM interactions that support a robust FM order. The DFT results imply that Cr-substitution results in strong and long-range FM pairwise interactions, making it an optimal candidate for the development of topological surface states. Overall, these results point to the fundamental chemical barriers that may prevent the development of robust ferromagnetism in Te-based topological insulators.  New routes that employ magnetic substitutions to discover Chern and axion insulating topological states are needed to carefully explore crystal structures that avoid these deleterious configurations.

\section{Acknowledgments} The work at the Ames National Laboratory was supported by the U.S. Department of Energy (USDOE), Office of Basic Energy Sciences, Division of Materials Sciences and Engineering.  Ames National Laboratory is operated for the USDOE by Iowa State University under Contract No. DE-AC02-07CH11358. A portion of this research used resources at the Spallation Neutron Source, which is a USDOE Office of Science User Facility operated by the Oak Ridge National Laboratory.

\section{AUthor Contribution}
R.~J.~M.~conceived and designed the experiments. D.~S., S.~P., and D.~C.~J.~synthesized the crystals and characterized them. F.~I., D.~M.~P.~, R.~J.~M., and D.~V.~conducted neutron scattering measurements. R.~J.~M., F.~I., and D.~V.~performed experimental data analysis and modeling. Y.~L.~and L.~K.~performed the DFT calculations. R.~J.~M.~prepared the figures. R.~J.~M., F.~I., L.~K.~and D.~V.~wrote the manuscript. All coauthors read and commented on the final manuscript.

\newpage

\bibliography{main_v2_PRX}

\end{document}